# RECONSTRUCTING THE ENERGY LANDSCAPE OF A DISTRIBUTION FROM MONTE CARLO SAMPLES[1]

BY QING ZHOU AND WING HUNG WONG

*University of California, Los Angeles and Stanford University*

Defining the energy function as the negative logarithm of the density, we explore the energy landscape of a distribution via the tree of sublevel sets of its energy. This tree represents the hierarchy among the connected components of the sublevel sets. We propose ways to annotate the tree so that it provides information on both topological and statistical aspects of the distribution, such as the local energy minima (local modes), their local domains and volumes, and the barriers between them. We develop a computational method to estimate the tree and reconstruct the energy landscape from Monte Carlo samples simulated at a wide energy range of a distribution. This method can be applied to any arbitrary distribution on a space with defined connectedness. We test the method on multimodal distributions and posterior distributions to show that our estimated trees are accurate compared to theoretical values. When used to perform Bayesian inference of DNA sequence segmentation, this approach reveals much more information than the standard approach based on marginal posterior distributions.

**1. Introduction.** The concept of a distribution is fundamental in many parts of modern science. In statistics we may model a set of observed data by assuming that they are sampled from a distribution specified up to some parameters, and then estimate the parameters based on the empirical data. Furthermore, if we use a Bayesian approach to statistical inference, then our knowledge of the parameters given the data is contained in the posterior distribution. In physics the Boltzmann distribution of a system in thermal

Received May 2008; revised July 2008.

[1]Supported in part by a faculty career development award at UCLA, NSF Grant DMS-08-05491 and NSF Grant DMS-05-05732. Part of the computation in this work was supported by NSF Grant CNS-0619926.

equilibrium at temperature $T$ is

$$p(\mathbf{x};T) = \frac{1}{Z(T)} \exp(-h(\mathbf{x})/T), \tag{1}$$

where $h(\mathbf{x})$ is the energy and $Z(T) = \int \exp(-h(\mathbf{x})/T)\, d\mathbf{x} < \infty$ is the normalization constant. Here the density (1) is defined with respect to a measure, so that discrete cases are covered by the use of counting measure. To unify terminology, we define the energy function of a distribution $f(\mathbf{x})$, which may be unnormalized, as

$$h(\mathbf{x}) = -\log f(\mathbf{x}). \tag{2}$$

One can view $f(\mathbf{x})$ as a Boltzmann distribution with energy $h(\mathbf{x})$ at temperature $T = 1$.

In many situations the distribution of interest is completely specified in the sense that we know how to compute $f(\mathbf{x})$ for any $\mathbf{x}$. However, in general, knowing the distribution in this way does not allow us to understand the information it embodies. To understand the nature of the distribution, we must seek answers to a multitude of questions, such as what is the expectation of a certain function $g(\mathbf{X})$ when a random variable $\mathbf{X}$ is drawn from this distribution, where is the mode of the distribution and how dispersed is the distribution around it, and are there multiple regions with high probabilities that are well separated in the sample space?

Before the development of modern numerical computing on digital computers, it was not possible to answer any of these questions except in very special cases, such as when $f(\mathbf{x})$ is a multivariate normal distribution. With the emergence of computers in the mid-twentieth century, physicists developed several Monte Carlo algorithms that allow the generation of samples from $f(\mathbf{x})$ numerically. In particular, Markov Chain based methods, such as the Metropolis–Hastings algorithm [Metropolis et al. (1953) and Hastings (1970)], can be applied to sample from a distribution in a very high dimensional space. Later, when statisticians adapted it to applications in Bayesian inference [Geman and Geman (1984); Tanner and Wong (1987); Gelfand and Smith (1990)], Monte Carlo sampling quickly became a popular means to extract information from a posterior distribution.

In principle, the availability of a large sample will allow us to understand the nature of the distribution by the use of standard data analysis tools. For example, the expectation of $g(\mathbf{X})$ can be estimated by the sample average and the distribution of $g(\mathbf{X})$ can be approximated by the corresponding histogram. Although powerful, such approaches can only provide limited information, as illustrated by the following example. The energy function $h(\mathbf{x})$ [equation (2)] in this example has seven local minima, as indicated by red numbers in Figure 1(A), and the global minimum is located at the origin. It is very hard to recover the seven modes from any projection of samples from this distribution. However, the topology of the interiors of various contours (sublevel sets) actually contains information about the local modes. At very low energy levels, the contours and their interiors form disconnected solid disks, each containing a local minimum. Then modes 2–4 become connected at energy level $h(\mathbf{x}) = 12$, and so do modes 5–7. At this level, the interiors of the contours become three disconnected regions. For an energy level $\geq 20$, the interior of the contour is completely connected, which links the three groups of modes together. Such information can be summarized by a tree of sublevel sets of the energy function [Figure 1(B)], in which terminal nodes represent the local minima and internal nodes give the energy at which the modes become connected (energy barriers). Such a tree was first studied by Hartigan (1975, 1981) in statistics and it is also related to the concept of a disconnectivity graph in chemical physics [Becker and Karplus (1997)]. Please see Section 2 for a rigorous definition.

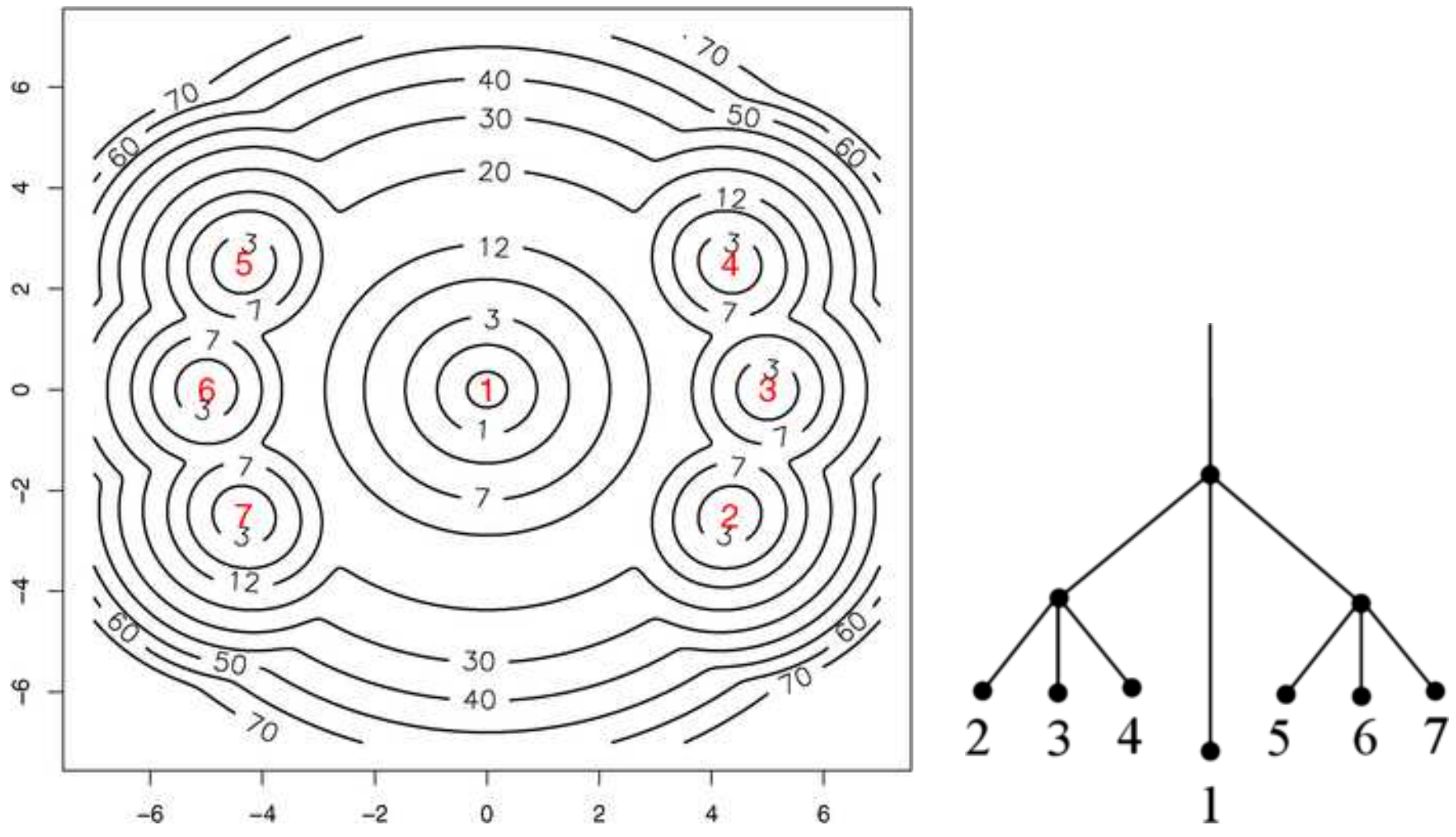

(A) (B)

Fig. 1. *An illustrative example.* (A) *The contour plot of the energy function of a 2-D distribution with seven modes indicated by red numbers. The black numbers are energy levels of the contours.* (B) *The tree of sublevel sets of the energy function.*

In this paper we propose a general method to estimate the tree of sublevel sets from Monte Carlo samples. We focus on the application of this method in understanding the energy landscape of a posterior distribution in Bayesian inference. This paper is organized into seven sections. Section 2 defines the tree of sublevel sets for a distribution. In Section 3 we develop an algorithm to estimate the tree and present related theoretical results. The method is tested on multimodal functions and posterior distributions in Section 4. A detailed application of this method in the Bayesian inference of DNA sequence segmentation is presented in Section 5. The paper is concluded with discussions in Section 6. Mathematical proofs are provided in the Appendix.

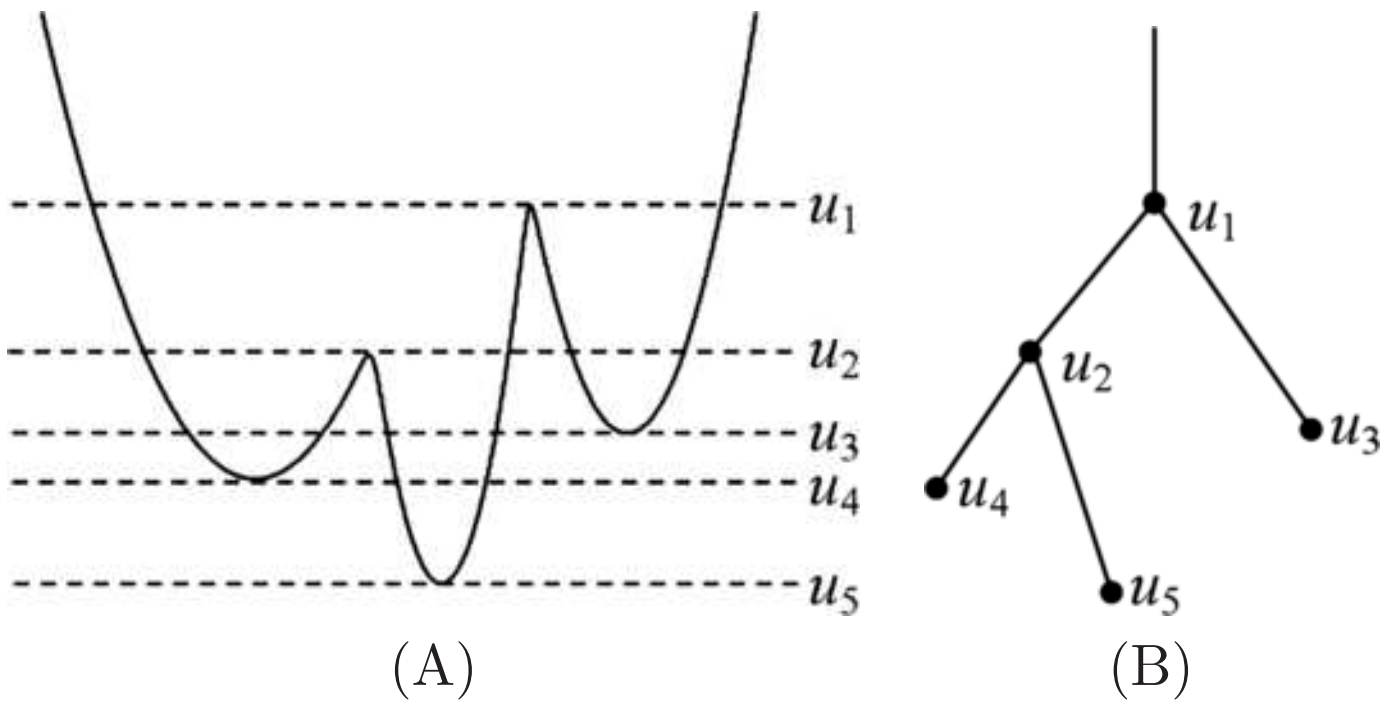

FIG. 2. *The definition of the tree of sublevel sets.* (A) *A hypothetical energy function.* (B) *The tree of sublevel sets of the energy function. In this tree internal nodes $u_1$ (root) and $u_2$ represent two energy barriers, while $u_3, u_4$ and $u_5$ are local minima.*

**2. The tree of sublevel sets.** Consider a continuous energy function $h(\mathbf{x})$ of a distribution (2) defined on a connected space $\mathcal{X}$ with (global) minimum $u_0$. For any $u > u_0$, suppose the sublevel set $A(u) = \{\mathbf{x} \mid h(\mathbf{x}) < u\}$ contains a finite number $K(u)$ of connected components $\{A_k(u) \mid k = 1, \dots, K(u)\}$. We may simply call them components if the meaning is clear from the context. As pointed out by Hartigan (1975, 1981), the collection $\mathcal{A} = \{A_k(u) \mid k = 1, \dots, K(u), u > u_0\}$ has a hierarchical structure: For any two sets $A_i(u_1)$ and $A_j(u_2)$ with $u_1 < u_2$, either $A_i(u_1) \subset A_j(u_2)$ or $A_i(u_1) \cap A_j(u_2) = \phi$. One can represent such hierarchy by a tree. The root of the tree is defined at the energy level $u_1 = \inf\{u | A(u) isconnectedand nonempty\}$. If $A(u)$ is connected for all $u > u_0$, then $u_1 = u_0$ and $A(u_1)$ is empty, which results in a terminal node. Otherwise, $A(u_1)$ is disconnected with $K(u_1) > 1$ components and it represents an internal node. We further define its $k$th child node at the energy level $u_{1k} = \inf\{u | A(u) \cap A_k(u_1)$ is connected and nonempty$\}$ for $k = 1, \dots, K(u_1)$. Recursively applying the above definition to each internal node defines the tree of sublevel sets. The leaves (terminal nodes) of the tree correspond to the local minima of $h(\mathbf{x})$ and the internal nodes correspond to the energy barriers that separate the minima. See Figure 2 for an illustration. Such a tree was also called the cluster tree in the context of clustering analysis [e.g., Stuezle (2003)].

To provide further information on the energy landscape, we propose to annotate the tree by local density of states. The density of states $\Omega(u)$ is a function of energy defined as the derivative of the volume of $A(u)$ with respect to $u$:

$$\Omega(u) = \frac{d}{du} \int \mathbf{1}(\mathbf{x} \in A(u))\, d\mathbf{x},$$

where $\mathbf{1}(\cdot)$ is the indicator function. By definition, the infinitesimal volume of the level set $\{\mathbf{x} \mid h(\mathbf{x}) \in [u-du, u)\}$ is $\Omega(u)\,du$. Similarly, for each component $A_k(u)$, we define

$$\Omega_k(u) = \frac{d}{du}\int \mathbf{1}(\mathbf{x} \in A_k(u))\,d\mathbf{x}$$

as the local density of states for $k = 1, \ldots, K(u)$. Obviously, $\Omega(u) = \sum_k \Omega_k(u)$. From local density of states one can readily compute many statistical properties of a local minimum at different temperatures. Suppose $\mathbf{x}_m$ is a local minimum of $h(\mathbf{x})$ whose parent on the tree of sublevel sets is $B$. Here we use $B$ to denote the node on the tree as well as its energy. For any $u > h(\mathbf{x}_m)$ there exists a unique integer $k \in \{1, \ldots, K(u)\}$ such that $\mathbf{x}_m \in A_k(u)$, and we denote this integer by $k(u, \mathbf{x}_m)$. Let $i = k(B, \mathbf{x}_m)$. Then $A_i(B)$ defines a unique local domain of the minimum before it reaches any energy barrier. For example, in Figure 2(B), the unique local domain of the node $u_4$ corresponds to the branch between the nodes $u_2$ and $u_4$ on the tree. It will be informative to compute expectations over such local domains with respect to the Boltzmann distribution (1). For instance, the probability of visiting $A_i(B)$ at temperature $T$, which is called the probability mass of the local minimum $\mathbf{x}_m$ hereafter, can be computed by

$$\begin{aligned} P(A_i(B); T) &= \int_{h(\mathbf{x}_m)}^{B} \frac{1}{Z(T)} \Omega_{k(u,\mathbf{x}_m)}(u) e^{-u/T}\,du \\ &= \frac{\int_{h(\mathbf{x}_m)}^{B} \Omega_{k(u,\mathbf{x}_m)}(u) e^{-u/T}\,du}{\int \Omega(u) e^{-u/T}\,du}, \end{aligned} \tag{3}$$

which only involves one-dimensional integrals. A similar formulation may be adapted to calculate marginal likelihood in Bayesian model selection as in Skilling (2006). Note that both the tree of sublevel sets and the local density of states are independent of temperature $T$ and intrinsically determined by the energy function. In statistics, this implies that once they are estimated for a distribution $f(\mathbf{x})$, we can use them to calculate expectations and describe the energy landscape for a tempered distribution, $[f(\mathbf{x})]^{1/T} \propto \exp(-h(\mathbf{x})/T)$, or a truncated one, $\exp(-(h(\mathbf{x}) \vee H)) \equiv \exp(-\max(h(\mathbf{x}), H))$.

**3. Estimation of the tree of sublevel sets.** Now we turn to the central question of this article: How to construct the tree of sublevel sets based on Monte Carlo samples from a distribution $f(\mathbf{x})$? It is practically impossible to obtain information about minima and barriers at high energy levels if we only have samples from the target distribution $f(\mathbf{x})$. To construct a reasonable estimate of the high energy portion of the tree, we need to generate samples from tempered versions of the distribution as used in parallel tempering [Geyer (1991)], that is, $f(\mathbf{x};T) \propto \exp(-h(\mathbf{x})/T)$, or from tempered-truncated versions as in the equi-energy sampler [Kou, Zhou and Wong (2006)], that is, $f(\mathbf{x};T,H) \propto \exp(-(h(\mathbf{x}) \vee H)/T)$. In what follows we assume that we have generated samples from a sequence of tempered or tempered-truncated distributions and for each sample we have recorded its energy $h(\mathbf{x})$. Estimating the tree is then equivalent to partitioning all the Monte Carlo samples into the components of various sublevel sets. Given $u_1 < u_2 < \cdots < u_M$, we define $M$ level sets (energy rings), $C^{(m)} = \{\mathbf{x} \mid h(\mathbf{x}) \in [u_{m-1}, u_m)\}$ for $m = 1, \ldots, M$, where $u_0 \equiv -\infty$.

3.1. *Connected components of level sets.* Let us use the example in Figure 1 to motivate our algorithm. In this example the energy values of mode 1 and the other six modes are 0 and 2, respectively. Let $u_m = m$ for $m = 1, \ldots, 70$. If we have partitioned the sublevel set $A(u_{m-1}) = \bigcup_{r \leq m-1} C^{(r)}$ into its connected components $\{A_j(u_{m-1}) \mid j = 1, \ldots, K(u_{m-1})\}$, then there exist only three possibilities to induce the partition of $A(u_m)$ from the components of $C^{(m)}$, denoted by $\{C_i^{(m)}\}$. First, $C_i^{(m)}$ is not connected to any components of $A(u_{m-1})$, which implies that it represents a terminal node (local minimum), such as the components containing minima 2 to 7 of the level set $C^{(3)} = \{\mathbf{x} \mid h(\mathbf{x}) \in [2, 3)\}$. Second, $C_i^{(m)}$ is connected to a single component $A_j(u_{m-1})$, $j \in \{1, \ldots, K(u_{m-1})\}$, such as any level set component on the branch between mode 1 and its parent (the barrier at energy $= 20$). Third, $C_i^{(m)}$ is connected to multiple components of $A(u_{m-1})$ and it corresponds to a barrier on the tree, for example, $C^{(21)} = \{\mathbf{x} \mid h(\mathbf{x}) \in [20, 21)\}$. Clearly, the components of level sets serve as the building blocks for an inductive construction of the tree.

3.2. *The main algorithm.* Define an empirical level set $\hat{C}^{(m)}$ and sublevel set $\hat{A}^{(m)}$ by the collections of samples in $C^{(m)}$ and in $A(u_m)$, respectively, for $u_1 < u_2 < \cdots < u_M$. A (connected) cluster of a set of samples generated in $D \subset \mathcal{X}$ is defined as the maximal subset of the samples in a connected component of $D$. Given a metric of the space, we employ single-linkage clustering (SLC) to partition an empirical level set into clusters. SLC recursively merges two closest subsets of samples according to the nearest neighbor distance (NND) between them. Define the maximum NND of a set of samples by the NND between the two subsets that are merged at the last step in the SLC. Based on NNDs and subset sizes, we develop statistical methods to identify clusters in $\mathbb{R}^p$ and in a discrete space with details given in Sections 3.4 and 5.3, respectively.

As illustrated in the previous subsection, we can construct the tree of sublevel sets by partitioning samples into clusters of empirical sublevel sets via a bottom-up induction. Thus, we call this method the bottom-up partition (BUP) algorithm, as outlined below.

1. Initialization: Perform SLC on $\hat{C}^{(1)}$ to obtain clusters $\{\hat{C}_k^{(1)}\}_1^{K_1}$ and the respective maximum NNDs of these clusters $\{d_k^{(1)}\}_1^{K_1}$. Let $\{\hat{A}_k^{(1)}\}_1^{K_1} = \{\hat{C}_k^{(1)}\}_1^{K_1}$ be the clusters of $\hat{A}^{(1)}$.
2. Induction: For $m = 2, \dots, M$:
   (a) Perform SLC on $\hat{C}^{(m)}$ to obtain clusters $\{\hat{C}_i^{(m)}\}_1^{K_m^*}$ and their maximum NNDs $\{r_i^{(m)}\}$;
   (b) Connect $\hat{C}_i^{(m)}$ and $\hat{A}_j^{(m-1)}$ if the NND between them is $\le \max(r_i^{(m)}, d_j^{(m-1)})$ for $i = 1, \dots, K_m^*$ and $j = 1, \dots, K_{m-1}$;
   (c) Merge the resulting connected clusters to obtain $\{\hat{A}_k^{(m)}\}_1^{K_m}$, the clusters of $\hat{A}^{(m)}$, and update their maximum NNDs $d_k^{(m)} = \max\{r_i^{(m)}, d_j^{(m-1)} | \hat{C}_i^{(m)}, \hat{A}_j^{(m-1)} \subset \hat{A}_k^{(m)}\}$ for $k = 1, \dots, K_m$.

If the distribution $f(\mathbf{x})$ is defined on a finite number of disconnected regions, this algorithm may build multiple trees, each for a connected component of the domain. Note that in step (2b) multiple $\hat{C}_i^{(m)}$'s may be connected to the same $\hat{A}_j^{(m-1)}$. This happens when different connected components of $C^{(m)}$ belong to the same component of $A^{(m)}$.

If $\hat{\Omega}_m$ is an estimated density of states for the $m$th level set, the algorithm also provides a simple way to approximate the local density of states:

$$\hat{\Omega}_{m,k} = n_k^{(m)}/n^{(m)} \cdot \hat{\Omega}_m \qquad \text{for } k = 1, \dots, K_m, \tag{4}$$

where $n^{(m)}$ and $n_k^{(m)}$ are the sample sizes of the level sets $\hat{C}^{(m)}$ and $\bigcup\{\hat{C}_i^{(m)} \mid \hat{C}_i^{(m)} \subset \hat{A}_k^{(m)}\}$, respectively. The estimation (4) follows immediately from the definition of density of states and the fact that samples in a level set are approximately uniform. In this work density of states is estimated by the iterative approach implemented in the equi-energy (EE) sampler [Kou, Zhou and Wong (2006), Section 4], which is also applicable to samples generated by parallel tempering.

3.3. *Theoretical considerations.* Define a discretized version of the tree of sublevel sets at discrete energy levels $\{u_m\}_{m=1}^M$ by the tree that represents the hierarchy among the collection $\{A_k(u_m) \mid k = 1, \dots, K(u_m), m = 1, \dots, M\}$. Intuitively, one may imagine to use $M$ horizontal lines at energy levels $u_1 < u_2 < \cdots < u_M$ to intersect the original tree. Each intersection represents a component $A_k(u_m)$ of the sublevel set $A(u_m)$ $(1 \le m \le M)$. Then we use a line segment to link $A_j(u_{m-1})$ to $A_k(u_m)$ if and only if $A_j(u_{m-1}) \subset A_k(u_m)$ for $m = 2, \dots, M$. The resulting graph is the discrete tree which can be viewed as an approximation to the original tree.

DEFINITION. Given $u_1 < u_2 < \cdots < u_M$, let $\hat{\mathcal{T}}_n = \{\hat{A}_k^{(m)} \mid k = 1, \ldots, K_m, m = 1, \ldots, M\}$ represent a tree constructed from an empirical sublevel set $\hat{A}^{(M)}$ of size $n$. We say that $\hat{\mathcal{T}}_n$ is consistent if the following statements hold in probability for all $m$ as $n \to \infty$:

(i) $K_m \to K(u_m)$;

(ii) $\sup_{\mathbf{x} \in A_k(u_m)} d(\mathbf{x}, \hat{A}_k^{(m)}) \to 0$, where $d(\mathbf{x}, A)$ is the minimal distance from $\mathbf{x}$ to the set $A$;

(iii) $\hat{A}_j^{(m-1)} \subset \hat{A}_k^{(m)}$ if and only if $A_j(u_{m-1}) \subset A_k(u_m)$.

The BUP algorithm will build a consistent tree if we have the following: (1) SLC on $\hat{C}^{(m)}$ can provide a consistent estimate of the components of $C^{(m)}$ in the sense of (i) and (ii) in the above definition; (2) $\hat{C}_i^{(m)}$ and $\hat{A}_j^{(m-1)}$ can be connected consistently in step (2b). Some theoretical considerations for the verification of these two conditions are provided.

LEMMA 1. *Let $f(\mathbf{x})$ be a continuous density on $\mathcal{X} \subset \mathbb{R}^p$. Suppose $D \subset \mathcal{X}$ is a compact subset with a connected interior $D^0$ and $f(\mathbf{x}) > 0$ $\forall \mathbf{x} \in D$. If an $f$-irreducible Markov chain $\{\mathbf{X}_t\}$ with invariant distribution $f$ is Harris recurrent, then the maximum NND of $\hat{D}_n = \{\mathbf{X}_t \mid \mathbf{X}_t \in D, t = 1, \ldots, n\}$ and $\sup_{\mathbf{x} \in D^0} d(\mathbf{x}, \hat{D}_n)$ converge to 0 almost surely as $n \to \infty$.*

The proof of this lemma is given in the Appendix. As discussed in Tierney (1994), most MCMC algorithms, such as irreducible Gibbs samplers and Metropolis algorithms, are Harris recurrent under mild conditions, to which Lemma 1 applies. Note that any nonempty $C^{(m)}$ has compact closure $\{\mathbf{x} \mid h(\mathbf{x}) \in [u_{m-1}, u_m]\}$, similarly for $A(u_m)$. If an empirical level set $\hat{C}^{(m)}$ is generated by multiple Harris recurrent Markov chains with invariant distributions strictly positive on $C^{(m)}$, such as tempered or truncated target distributions, then condition (1) will be satisfied if the distance between any two components of $C^{(m)}$ is positive.

LEMMA 2. *Suppose $f(\mathbf{x})$ and $D$ satisfy the same conditions in Lemma 1. A random sample $\{\mathbf{X}_i\}$ of size $n$ is drawn from $f$ and SLC is performed on $\hat{D}_n = \{\mathbf{X}_i \in D\}$ with a distance threshold $\rho/n$. For some $\rho$, there exists a big cluster that includes a positive fraction of $\hat{D}_n$ and passes within $\varepsilon_n$ of every element of $\hat{D}_n$, and every other cluster has diameter (maximum within-cluster distance) $< \varepsilon_n$, where $\varepsilon_n \to 0$ in probability as $n \to \infty$.*

Lemma 2 is a mild modification of Theorem 1 in Hartigan (1981). If $C_i^{(m)}$ and $A_j^{(m-1)}$ are connected and each has a positive volume, then there is a distance threshold $\rho/n$ with which SLC produces a big cluster that contains fractions of samples in both sets, while every other cluster is arbitrarily small. This implies that $\rho/n < \max(r_i^{(m)}, d_j^{(m-1)})$ and $P(R_n < \rho/n) \to 1$ as $n \to \infty$, where $R_n$ is the NND between $\hat{C}_i^{(m)}$ and $\hat{A}_j^{(m-1)}$. Thus, condition (2) for the consistence of the algorithm will be satisfied with an i.i.d. sample as the input. If $\{\mathbf{X}_t\}$ is a Markov chain as stated in Lemma 1, one can apply the BUP algorithm to a subsequence $\{\mathbf{X}_{t_i} \mid i = 1, \dots, n\}$ with $(t_{i+1} - t_i)$ sufficiently large such that this subsequence behaves like an i.i.d. sample from $f$.

3.4. *Clustering level sets in* $\mathbb{R}^p$. Given that samples in a level set are approximately uniform if $(u_m - u_{m-1})$ is small, we consider the following results, which are proved in the Appendix.

LEMMA 3. *Let $h(\mathbf{x})$ be a continuous function in $\mathbb{R}^p$. Suppose a connected level set $C = \{\mathbf{x} \mid h(\mathbf{x}) \in [u - \Delta u, u)\}$ has a finite volume $V_C > 0$ and a random sample of size $n$, $\{\mathbf{X}_1, \dots, \mathbf{X}_n\}$, is drawn uniformly on $C$. Let $n \to \infty$.*

(1) *Denote by $r_i$ the NND between $\mathbf{X}_i$ and the other sample points. Then $nr_i^p/V_C$ follows an identical exponential distribution (denote its mean by $\theta$) and is independent of $nr_j^p/V_C$ $(j \neq i)$.*

(2) *Denote by $d$ the NND between a finite subset $\mathbf{X}^* = \{\mathbf{X}_{i_1}, \dots, \mathbf{X}_{i_k}\}$ and the other sample points. Then given $\mathbf{X}^*$, $nd^p/V_C$ follows an exponential distribution with mean $\theta/\beta (\beta > 1)$.*

This lemma suggests that the NNDs in the SLC on an i.i.d. uniform sample decay exponentially fast in the order of $(V_C/n)^{1/p}$ for points within a connected component, which become significantly smaller than between-component distances (BCD) when the sample size is large. Thus, one may treat BCDs as outliers in an exponential sample and develop methods to detect them. Suppose that a random sample $\{Y_1, \dots, Y_n\}$ is drawn from $\mathrm{Exp}(\theta)$ with mean $\theta$ and that the largest $k$ observations are missing. Denote the observed order statistics by $y_{(1)} < \cdots < y_{(n-k)}$. Then the observed data likelihood is

$$L(\theta | y_{(1)}, \dots, y_{(n-k)}) = [\exp(-y_{(n-k)}/\theta)]^k \prod_{i=1}^{n-k} \theta^{-1} \exp(-y_{(i)}/\theta),$$

which leads to the MLE of $\theta$:

$$\hat{\theta}_k = \frac{\sum_{i=1}^{n-k} y_{(i)} + k y_{(n-k)}}{n-k}.$$

Let $y_i = nr_i^p$ with $r_i$ $(i = 1, \dots, n)$ the NNDs in the SLC of an i.i.d sample of $(n+1)$ points from a level set $C$. Suppose that $C$ has $K+1$ connected components and the NNDs among them are $d_{(K)} \geq \cdots \geq d_{(1)} > 0$. As $n \to \infty$, we have approximately

$$\hat{\theta}_k \to \begin{cases} \theta + \sum_{i=1}^{K-k} d_{(i)}^p + k d_{(K-k)}^p, & \text{for } 0 \leq k \leq K-1, \\ \theta, & \text{for } k \geq K. \end{cases}$$

This suggests that one may estimate $K$ by the value of $k$ from which $\hat{\theta}_k$ starts to stabilize. In order to choose a more interpretable threshold, we define $\hat{P}_k \propto 1/\hat{\theta}_k$ for $k = 0, 1, \dots, K_{\max} - 1$ and normalize $\hat{P}_k$ such that $\sum_k \hat{P}_k = 1$, where $K_{\max} \gg K$ is a pre-determined maximal number of components of a level set. Consequently, $\hat{P}_k$ increases to a level slightly above $1/K_{\max}$ with the increase of $k$, as illustrated in Figure 3. For this particular level set with seven components, $\hat{P}_0, \dots, \hat{P}_5 \ll 1/K_{\max}$, while the other $\hat{P}_k$'s are all very close to and slightly above $1/K_{\max}$. Such a pattern allows us to define two bounds for the number of components, $K_b = 1 + \min\{k \mid \hat{P}_k > \delta_b \cdot 1/K_{\max}\}$ for $b = L, H$, in which $0 < \delta_L < \delta_H < 1$. From the SLC of an empirical level set $\hat{C}^{(m)}$, we first obtain $K_L$ clusters. If we gradually increase the number of clusters from $K_L$ to $K_H$, $(K_H - K_L)$ clusters will be split sequentially. We discard a resulting daughter cluster from a split if it contains $\leq N_{\min}$ points. The remaining ones form the clusters $\{\hat{C}_i^{(m)}\}_1^{K_m^*}$ for the level set. Such a cluster either contains more than $N_{\min}$ points or its NND is among the largest $K_L - 1$. This procedure rules out those small and often false clusters with moderate between-cluster distances. From our empirical studies, this approach seems to work very well even for samples generated from a Markov chain when the sample size is reasonably large.

3.5. *Practical issues.* First, the complexity of the BUP algorithm is dominated by SLC of empirical level sets. If the samples are generated by an MCMC method, we typically resample without replacement about 20% of the samples as the input. This can reduce the computation greatly without degrading the performance. We divide the samples into enough level sets such that the size of each set is in the order of 5K to 10K. Second, the default values for the parameters in level set clustering are specified as $\delta_L = 0.5$, $\delta_H = 0.95$, $K_{\max} = 100$ and $N_{\min} = 50$. These values are used in all the examples presented in this article. We note that, for a level set of size 5K or more, the performance of the algorithm is not sensitive to the choice of these parameters. For some distributions, the algorithm tends to underestimate the number of components $(K_m^*)$ when a level set is close to an energy barrier since the between-cluster distance tends to be small.

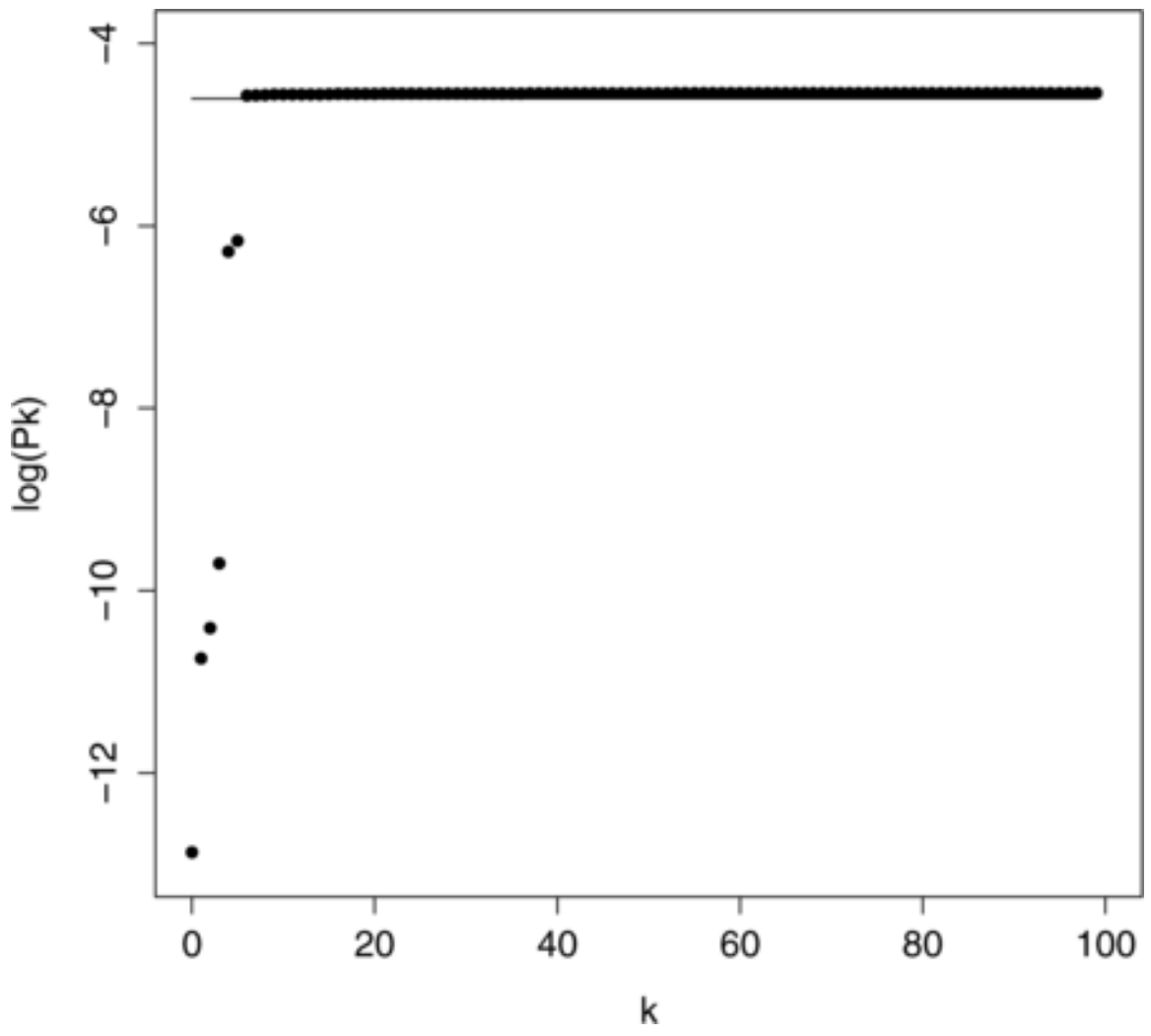

FIG. 3. *The* $\log(\hat{P}_k)$ *(solid dots,* $k = 0, \ldots, 99$*) of an empirical level set constructed from MCMC samples of a 5-D mixture normal distribution. This level set has seven components. The height of the horizontal solid line* $= \log(1/K_{\max}) = -\log(100)$.

Considering that another reasonable upper bound for the number of components of $C^{(m)}$ is $K_{m-1}$, the estimated number of components of $A(u_{m-1})$, if no new minima occur, we modify our upper bound to be $\max(K_H, K_{m-1})$, where $K_H$ is the original bound defined by $\delta_H$. Finally, a more efficient way to determine whether $\hat{C}_i^{(m)}$ and $\hat{A}_j^{(m-1)}$ should be connected in step (2b) of the BUP algorithm is to sequentially compute the distances between $\hat{C}_i^{(m)}$ and different level sets in $\hat{A}_j^{(m-1)}$ in the descending order of their energy. We stop the computation once we identify a level set whose NND to $\hat{C}_i^{(m)}$ is $\leq \max(r_i^{(m)}, d_j^{(m-1)})$.

The BUP algorithm may take as input from a variety of Monte Carlo methods besides the EE sampler and parallel tempering. The multicanonical sampling [Berg and Neuhaus (1991)] and related methods [Hesselbo and Stinchcombe (1995); Wang and Landau (2001); Liang (2005, 2007); Atchadé and Liu (2006); Liang, Liu and Carroll (2007)] can generate samples at various energy levels and estimate density of states. The outputs from these algorithms should be suitable for our method to estimate the tree of sublevel sets as well. With slight modifications, the nested sampling proposed by Skilling (2006) may be another candidate sampler to produce inputs for the BUP algorithm.

**4. Examples in a continuous space.** To demonstrate the use of the BUP algorithm in constructing the tree of sublevel sets, we test it on posterior distributions given multivariate $t$ data and a multimodal distribution with many local modes.

4.1. *Posterior inference from multivariate $t$ data.* Suppose we have observed a random sample $\mathbf{Y} = \{\mathbf{y}_1, \ldots, \mathbf{y}_n\}$ from a multivariate $t$ distribution $t_\nu(\boldsymbol{\mu}, \boldsymbol{\Sigma})$ with known degree of freedom and scale matrix, $\nu = 5$ and $\boldsymbol{\Sigma} = \mathbf{I}_p$, where $p$ is the dimensionality of the distribution. With a flat prior, our goal is to make inference on the location parameter $\boldsymbol{\mu}$ from its posterior distribution given the data $\mathbf{Y}$. In this case, the energy function of the posterior distribution is

$$h(\boldsymbol{\mu}) = -\log[P(\boldsymbol{\mu}|\mathbf{Y})] = \frac{\nu + p}{2} \sum_{i=1}^{n} \log\left[1 + \frac{1}{\nu}\|\mathbf{y}_i - \boldsymbol{\mu}\|^2\right], \tag{5}$$

up to an additive constant. Since multimodality of a $t$-likelihood occurs with an appreciable chance for a small sample size, we design an observed data matrix ($n = 6$) as

$$\begin{bmatrix} \mathbf{y}_1 \\ \mathbf{y}_2 \\ \mathbf{y}_3 \\ \mathbf{y}_4 \\ \mathbf{y}_5 \\ \mathbf{y}_6 \end{bmatrix} = \begin{bmatrix} A & A & a_1 & a_1 & 0 & 0 \\ A & A & 0 & 0 & a_1 & a_1 \\ a_2 & a_2 & A & A & 0 & 0 \\ 0 & 0 & A & A & a_2 & a_2 \\ a_3 & a_3 & 0 & 0 & A & A \\ 0 & 0 & a_3 & a_3 & A & A \end{bmatrix}, \tag{6}$$

where $A \gg a_j > 0$ ($j = 1, 2, 3$). Note the heterogeneity of the observed data in the sense that they form three sub-groups, each composed of two data points, in the 6-D space.

We first set $A = 40$ and $a_1 = a_2 = a_3 = 4$ so that the data set is composed of three symmetric pairs of points. To set up the energy ladder for the EE sampler, we did the following pilot study on the posterior distribution. We randomly generated 100 points from the hyper-cube defined by the boundary values of the observed data, that is, $[0, 40]^6$, evaluated their energy (5), and chose the maximum as the upper bound for the energy ladder. We then performed a few runs of gradient-based minimization to obtain the energy values of some local minima, and set the lower bound for the energy ladder as the smallest local minimum minus 3. In this way, the energy ladder was set geometrically between $[166, 220]$ and the temperature ladder between $[0.2, 4]$. Note that a higher temperature for this target distribution would cause an improper posterior. The combination of the energy and temperature ladders allowed the EE sampler to generate samples in a wide energy range from local minima to high energy barriers. We utilized 10 chains, each generating 200K samples, and resampled 20% of the samples to estimate the tree with $M = 50$ level sets. This computation was repeated 10 times, each with an independent input of EE samples. The topology of the constructed tree, exactly identical among different EE samples, is shown in Figure 4(A) with critical energy values reported in the left column of Table 1. The estimated tree of sublevel sets is composed of three long branches for a wide range of energy levels from $B_4 = 197.5$ to $B_i = 171.0$ $(i = 1, 2, 3)$. Further down the energy level, each of these branches splits into two symmetric local minima. Each local minimum $M_i$ is located near a data point $\mathbf{y}_i$ for $i = 1, \dots, 6$. To verify this constructed tree, we utilized a gradient-based local search from many random initial values, including the six data points to identify local minima, as reported in the column "Approx." in Table 1, which gave us exactly the same six local modes as identified on the tree. Furthermore, we approximated the energy barriers by finding the maximal energy along the line segment between every pair of local minima. These approximated barriers are all very close to what we have obtained on the tree (Table 1).

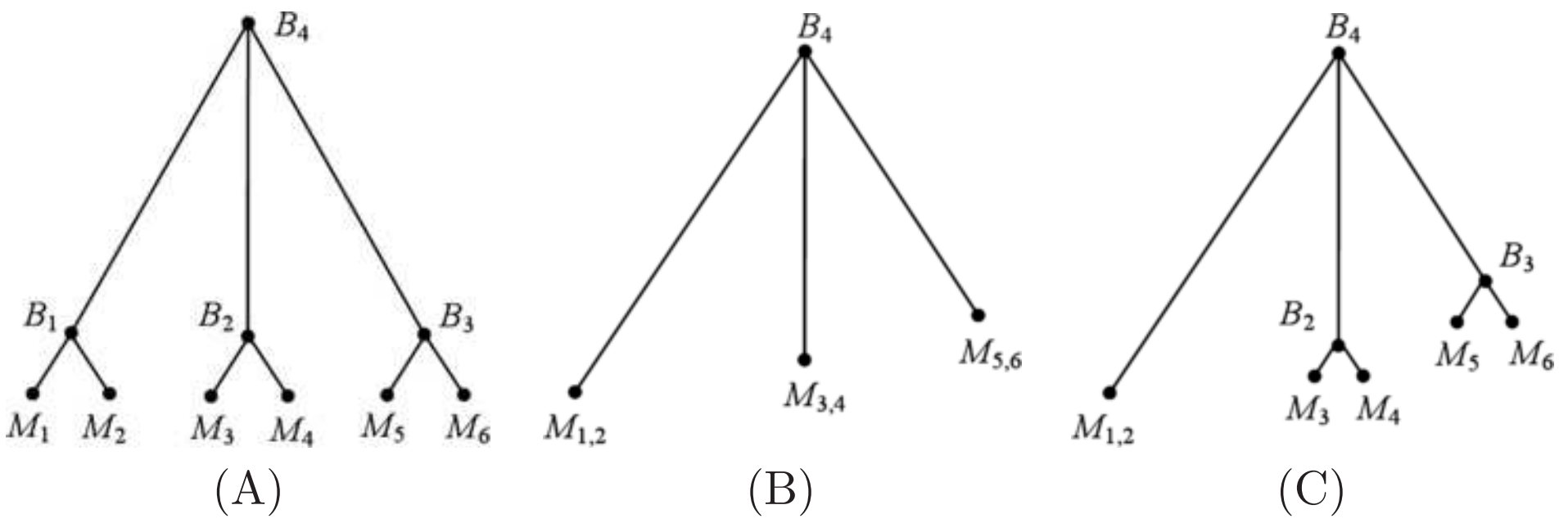

FIG. 4. *Estimated trees of sublevel sets for the posterior distributions given the t data:* (A) *symmetric data set,* (B) *asymmetric data set without local sampling, and* (C) *asymmetric data set with local sampling. Nodes are indexed in correspondence among the three trees.*

TABLE 1
*Critical energy estimation for the posterior distributions given the t data*

| | Symmetric | | Asymmetric | |
|---|---|---|---|---|
| **Node** | **Estimate (s.e.)** | **Approx.** | **Estimate (s.e.)** | **Approx.** |
| $M_1$ | 169.20 (0.015) | 169.18 | 162.34 (0.003) | 162.33 |
| $M_2$ | 169.20 (0.007) | 169.18 | — | |
| $M_3$ | 169.20 (0.006) | 169.18 | 166.59 (0.008) | 166.57 |
| $M_4$ | 169.20 (0.009) | 169.18 | 166.58 (0.009) | 166.57 |
| $M_5$ | 169.20 (0.006) | 169.18 | 169.62 (0.007) | 169.60 |
| $M_6$ | 169.21 (0.012) | 169.18 | 169.62 (0.008) | 169.60 |
| $B_1$ | 171.0 (0.075) | 170.9 | — | |
| $B_2$ | 171.0 (0.084) | 170.9 | 167.0 (0.018) | 167.0 |
| $B_3$ | 171.0 (0.074) | 170.9 | 171.4 (0.040) | 171.3 |
| $B_4$ | 197.5 (0.273) | 198.5 | 198.5 (0.893) | 200.5 |

From the estimated local density of states, we computed that the probability mass of each local mode (3) is only 0.015, while the total probability of the three long branches between $B_4$ and $B_i$ $(i = 1, 2, 3)$ is around 0.91. This suggests that the three within-component averages of the empirical sublevel set $\hat{A}(B_4)$ form a good representation of the posterior distribution [Figure 4(A)]. It is definitely better than the overall posterior mean which is located outside of any high probability region. It is also more appropriate than the six local modes which seem to "overfit" the small highest probability regions with posterior probability $< 0.1$. Clearly, the tree of sublevel sets provides much more information to interpret the posterior distribution than marginal averages and posterior modes.

Next we reset $a_1 = 2$ and $a_2 = 3$ to obtain an asymmetric data matrix in (6). After a similar pilot study on the energy function (5) to set the energy ladder in $[160, 220]$, we applied the EE sampler followed by the BUP algorithm with exactly the same parameter settings for 10 independent runs. There are only three branches on each of the constructed trees [Figure 4(B)], with a global minimum $(M_{1,2})$ near the center of $\mathbf{y}_1$ and $\mathbf{y}_2$. The other two local minima $M_{i,j}[(i, j) = (3, 4), (5, 6)]$ show larger variability: they are close to either $\mathbf{y}_i$ or $\mathbf{y}_j$. This implies that we may need to generate more samples on these two branches to refine our estimates. If we naively increased the sample size in the EE sampler, it would cause a much heavier computational burden on tree construction without any obvious improvement for the problem, since the local density of states on the branch $[M_{1,2}, B_4]$ is exponentially larger than those on the other two branches. However, with a coarsely estimated tree, one can design a more efficient way to refine the local sampling of a branch. Given an energy level $H^*$ between $B_4$ and $M_k$ for $k = (3, 4)$ or $(5, 6)$, we wish to restrict the EE sampler to the connected component $A_k(H^*)$ that contains the minimum $M_k$. This can be achieved by defining a modified energy function,

$$h_k(\boldsymbol{\mu}) = \begin{cases} h(\boldsymbol{\mu}), & \text{if } \boldsymbol{\mu} \in A(H^*) \cap B(M_k, d_k), \\ \infty, & \text{otherwise,} \end{cases}$$

where $B(M_k, d_k)$ is the ball centered at $M_k$ with radius $d_k$. Based on the coarse tree, we chose $d_k$ and $H^*$ such that $A(H^*) \cap B(M_k, d_k) = A_k(H^*)$. Then we performed EE sampling from the two modified local energy functions for $M_{3,4}$ and $M_{5,6}$, respectively, with 5 chains of 200K samples. The energy ladders were set between $(M_k - 2)$ and $H^* = (0.3M_k + 0.7B_4)$. Here we use $M_k$ to denote as well the energy of the local minimum. Other parameters for sampling and tree construction were identical to the previous calculations. This refined tree estimation with local sampling was performed for 10 independent runs based on the estimated coarse trees. The refined tree is shown in Figure 4(C) with critical energy values reported in the right column of Table 1, which are consistent with gradient-based approximations.

Now one sees that two small minima have been recovered on each of the refined branches. Note that by the volume of $A_k(H^*)$, one can renormalize the local density of states estimated from local sampling to obtain its corresponding value on the coarse tree, and thus perform related probabilistic calculation. For instance, the probability masses are estimated to around $1.2\times 10^{-4}$ for $M_3$ and $M_4$, and around $1.7\times 10^{-4}$ for $M_5$ and $M_6$. This example demonstrated the flexibility and power to manipulate posterior sampling with the aid of the tree of sublevel sets.

4.2. *The Rastrigin function.* We further test our method on a distribution with a large number of local modes. Let $\mathbf{x} = [x_1, \dots, x_p]$. The Rastrigin function [Gordon and Whitley (1993)] is defined as

$$h(\mathbf{x}) = \sum_{i=1}^{p} x_i^2 + A\left[p - \sum_{i=1}^{p} \cos(\pi x_i)\right], \tag{7}$$

where $A$ is a positive constant and $p$ is the dimensionality of the variable $\mathbf{x}$. This is one of the benchmark functions used to test a global optimization algorithm such as the genetic algorithm [Holland (1975)]. Although closely related, our purpose of constructing the tree of sublevel sets is more challenging than global optimization. We take $A = 2$ and $p = 4$ in (7) to obtain an energy function with $3^4 = 81$ local minima formed by all the elements of the product set $\{-1.805, 0, +1.805\}^4$. These minima have five distinct energy values shown in the theoretical tree [Figure 5(A)], dependent on the combinations of their coordinates. Correspondingly, we group them into five layers so that the $j$th layer contains local minima whose coordinates are composed

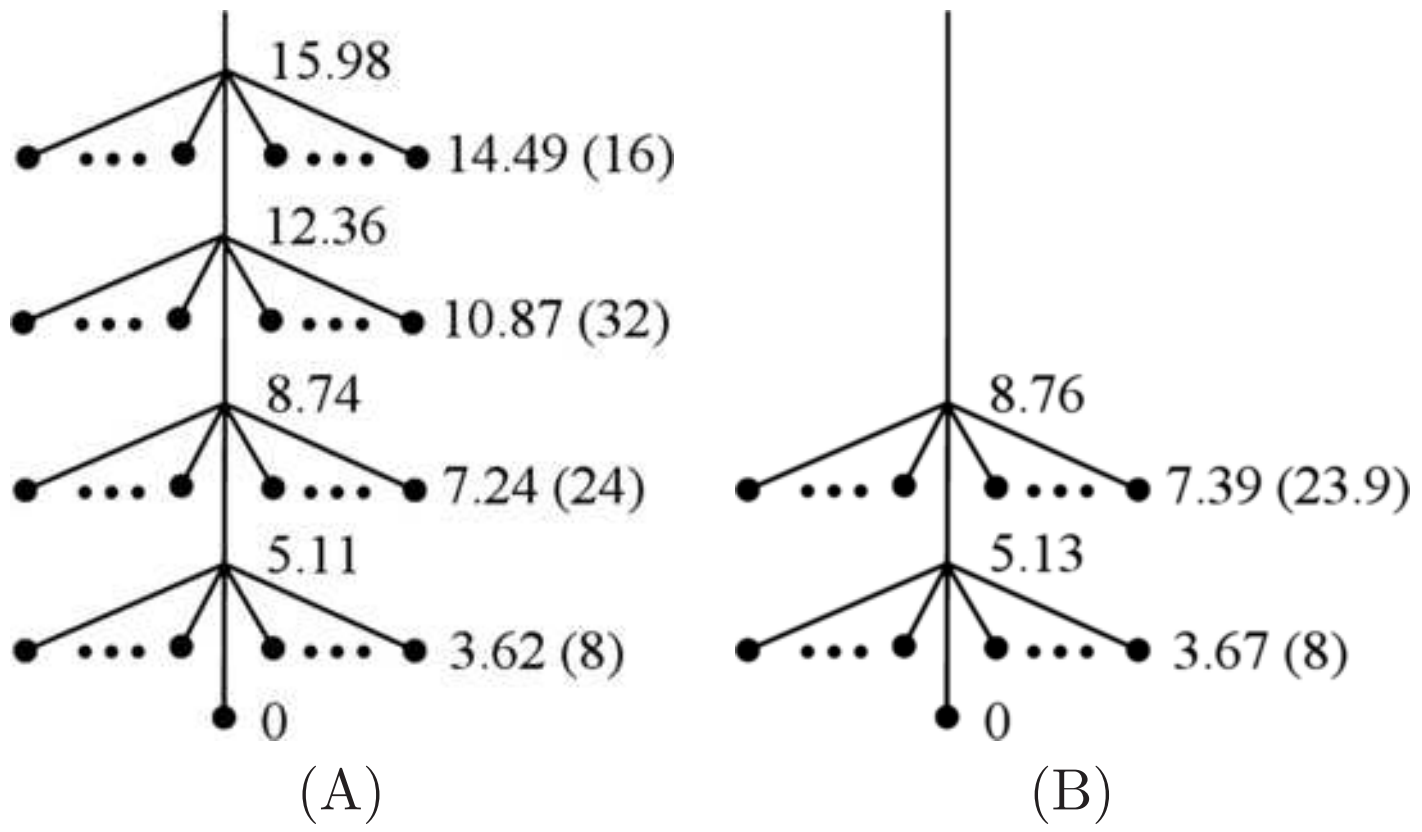

FIG. 5. *The trees of sublevel sets of the Rastrigin function.* (A) *Theoretical tree;* (B) *Estimated tree. The critical energy values are labeled on the trees and the number of minima in each layer is indicated in the parentheses.*

Table 2
*Local minima and barriers of the 4-D Rastrigin function*

| | Layer | 1 | 2 | 3 | 4 | 5 |
|---|---|---|---|---|---|---|
| A | Count | 1 | 8 | 24 | 32 | 16 |
| | Energy | 0 | 3.62 | 7.24 | 10.87 | 14.49 |
| | Barrier | — | 5.11 | 8.74 | 12.36 | 15.98 |
| B | Count | 1 (0) | 8 (0) | 12.1 (2.96) | 0 (0) | 0 (0) |
| | Energy | 0.005 (0.003) | 3.70 (0.070) | 7.62 (0.107) | — | — |
| | Barrier | — | 5.13 (0.085) | 8.54 (0.318) | — | — |
| C | Count | 1 (0) | 8 (0) | 23.9 (0.32) | 0 (0) | 0 (0) |
| | Energy | 0.005 (0.003) | 3.67 (0.023) | 7.39 (0.074) | — | — |
| | Barrier | — | 5.13 (0.085) | 8.76 (0.151) | — | — |

Note: "Count" and "Energy" refer to the number and the energy of the local minima in a layer, respectively. "Barrier" is the energy barrier between the current and the previous layers [see Figure 5(A)]. Theoretical values and estimated values without/with linear interpolation are given in panels A, B and C, respectively. The standard errors of estimates in the same layer from 10 independent EE samples are given in the parentheses.

of $(5-j)$ zeros and $(j-1)$ $\pm 1.805$, for $j=1,\dots,5$. The theoretical values of the local minima and energy barriers are given in Table 2 panel A.

We applied the EE sampler to this energy function with 20 chains, in which the energy was truncated evenly between $[0,19]$ and the temperature was fixed at $T=0.5$. Thus, the target distribution of the $k$th chain was $f_k(\mathbf{x}) \propto \exp[-(h(\mathbf{x}) \vee k)/T]$ for $k=0,\dots,19$ and $T=0.5$. We generated 100K samples from each chain and resampled 20% of them to construct the tree with $M=50$ level sets. This whole process was repeated 10 times independently. As reported in panel B of Table 2, for all the 10 independent inputs of EE samples, the BUP algorithm identified all the 9 minima in the first two layers and about half of those in the third layer. It also detected unambiguously the energy barriers associated with these three layers of minima.

There might be two reasons why our method failed to identify some high-energy local modes in layer three and beyond. First, the EE sampler did not visit them because of the tiny probability associated with these modes. Even for a truncated energy function, the local density of states of such a mode may be much smaller than that of the connected component which contains low-energy modes, and thus, the EE sampler has almost no chance to explore them. For example, the ratio of the local density of states of a mode in the fourth layer at energy $u=10.9$ over that of the middle main branch which connects to lower energy nodes on the tree [Figure 5(A)] is approximately $3\times 10^{-5}$. Intuitively, one may think of the main branch as much "thicker" than the leaves of the same height. Second, it is possible that the EE sampler visited some of these modes, but with insufficient samples they were not identified by the BUP algorithm as clusters of a level set. Recall that a detected cluster with a moderate NND (ranked between $K_L$ and $K_H$) must have at least $N_{\min} = 50$ samples, which might be too stringent for a small high-energy mode. This motivated us to use linear interpolation of the energy function to enhance the sensitivity in identifying clusters of a level set. Given $K_L$ and $K_H$, consider splitting a cluster in-between as described in Section 3.4. Suppose a resulting daughter cluster $D$ has less than $N_{\min}$ points. We define for $D$ a pair of points in the current empirical level set by $(\mathbf{x}^*, \mathbf{y}^*) = \arg\min_{\mathbf{x}\in D, \mathbf{y}\notin D} d(\mathbf{x}, \mathbf{y})$, which determine the NND for agglomerating $D$ with its closest cluster (sister cluster). Note that $\mathbf{x}^*$ and $\mathbf{y}^*$ can be obtained along with single-linkage clustering and no additional computation is needed. Then we evaluate the energy values of 100 points evenly distributed along the line segment between $\mathbf{x}^*$ and $\mathbf{y}^*$. The maximal energy of these interpolated points serves as an approximation to the barrier between $\mathbf{x}^*$ and $\mathbf{y}^*$ or between $D$ and its nearest neighbor cluster. Consequently, $D$ will not be discarded if the maximal interpolated energy is higher than the upper energy bound of the current level set.

When we applied the BUP algorithm with this linear interpolation to the same ten sets of EE samples from the Rastrigin function, it identified almost all the 33 local minima in the first three layers with reduced biases and standard errors (Table 2 panel C). For nine sets of samples, the method identified all the 33 local modes exactly. For the other one set, it missed one local mode in the third layer. All the constructed trees have a similar topology as shown in Figure 5(B). Although it failed to recover the tree topology for layers four and five, our approach still provided a global understanding of the energy landscape of this distribution since any statistical property of the distribution is unlikely to be affected, in practice, by ignoring those high-energy modes of tiny probabilities.

**5. Bayesian segmentation of DNA sequences.** To account for the heterogeneity of DNA sequences, statistical models have been proposed to segment a DNA sequence into pieces with more homogeneous nucleotide compositions [e.g., Liu and Lawrence (1999); Boys and Henderson (2004); Keith (2006)]. Bayesian inference on these models is performed via posterior sampling of segmentations (or change points). In this study we adopt the model of Liu and Lawrence (1999) to sample from a posterior distribution of change points. Then we apply the BUP algorithm to reconstruct the energy landscape of the posterior distribution which is expected to be quite complicated.

5.1. *A Bayesian model.* Denote a DNA sequence of length $L$ by $\mathbf{Y} = [y_1, \dots, y_L] \equiv y_{1:L}$, where $y_l \in \{a, c, g, t\}$ for $l = 1, \dots, L$. Assume that there exist at most $N$ change points. Let $\mathbf{Z} = [z_1, \ldots, z_p]$ $(0 \le p \le N)$ denote the locations of change points, which segment the sequence into $p+1$ pieces. Note that $p = 0$ implies that the whole sequence is in one segment. We assume that the prior distribution for the number of change points $p$ is uniform on $\{0, 1, \ldots, N\}$. Given $p$, we assume that every possible placement of the $p$ change points is equally likely in the prior. Within a segment defined by the change points, the nucleotides follow a multinomial distribution. We are interested in the posterior distribution of the locations of the change points $\mathbf{Z}$ given the sequence data $\mathbf{Y}$,

$$P(\mathbf{Z}|\mathbf{Y}) \propto \pi(p)\pi(\mathbf{Z}|p)P(\mathbf{Y}|\mathbf{Z}), \tag{8}$$

where $\pi(\cdot)$ is used as a generic notation for prior distributions. Note that under a conjugate Dirichlet prior for the multinomial distribution in a segment, $P(\mathbf{Y}|\mathbf{Z})$ in (8) can be computed exactly by ratios and products of gamma functions.

5.2. *Posterior sampling.* Exact sampling from the posterior distribution (8) can be achieved via dynamic programming as used in Liu and Lawrence (1999). This approach first samples from the marginal posterior distribution of $p$ (the number of change points) and then samples sequentially all the $p$ change points from $z_p$ to $z_1$. The key to this exact sampling is a recursion on the conditional probability of observing a partial sequence $y_{1:l}$ given that it has $k$ change points $p_{1:l} = k$,

$$\begin{aligned} P(y_{1:l}|p_{1:l} = k) = \sum_{z_k=k+1}^{l} & P(y_{1:z_k-1}|p_{1:z_{k-1}} = k-1) \\ & \times P(y_{z_k:l}|p_{z_k:l} = 0)\pi(z_k|p_{1:l} = k), \end{aligned} \tag{9}$$

where $\pi(z_k|p_{1:l} = k)$ is the conditional prior probability to place the $k$th change point at $z_k$, given that there are $k$ change points between $y_1$ and $y_l$. We pre-compute the probability of every subsequence $y_{i:l}$ $(1 \le i \le l \le L)$, given that it is generated from a multinomial distribution, that is, $P(y_{i:l}|p_{i:l} = 0)$, which can be calculated in closed-form based on gamma functions as we mentioned before. Then recursive forward summation (9) is applied to compute $P(y_{1:l}|p_{1:l} = k)$ for $k = 0, \ldots, N$ and $l = 1, \ldots, L$. After all the summations, we sample the number of change points $p = p_{1:L}$ from $P(p_{1:L} = k|\mathbf{Y}) \propto \pi(k)P(y_{1:L}|p_{1:L} = k)$. Given $p$, one can sequentially impute the change points $z_p, \ldots, z_1$ based on the additive terms in the summation (9). Please refer to Liu and Lawrence (1999) for more details.

We define the energy function of $\mathbf{Z}$ (a set of change points or a segmentation) by

$$h(\mathbf{Z}) = -\log P(\mathbf{Z}|\mathbf{Y}) \tag{10}$$

and a tempered distribution at temperature $T$ by

$$P(\mathbf{Z}|\mathbf{Y};T) \propto \exp(-h(\mathbf{Z})/T) = [P(\mathbf{Z}|\mathbf{Y})]^{1/T}.$$

It is easy to see that one can use a similar dynamic programming method to sample from $P(\mathbf{Z}|\mathbf{Y};T)$ for any $T > 0$. Our strategy for this problem is to generate (independent) samples at various temperatures and use them to reconstruct the energy landscape of the posterior distribution.

5.3. *Metric and clustering.* The space of change points is discrete in nature. We need to define connectedness in this space based on some metric. A natural choice of a metric between two sets of change points (segmentations) is the number of sequence positions that are partitioned into distinct segments. Write the segments defined by $\mathbf{Z} = [z_1, \ldots, z_p]$ as $\{[z_{k-1}, z_k) \mid k = 1, \ldots, p+1\}$, where $z_0 \equiv 1$ and $z_{p+1} \equiv L + 1$, and the segments by $\mathbf{X} = [x_1, \ldots, x_q]$ as $\{[x_{j-1}, x_j) \mid j = 1, \ldots, q+1\}$ similarly. For a one-to-one map $g$ from a subset of $\{1, \ldots, p+1\}$ into $\{1, \ldots, q+1\}$, the total number of common sequence positions between all pairs of mapped segments is

$$S_g(\mathbf{Z}, \mathbf{X}) = \sum_{k:g(k)\neq\phi} |[z_{k-1}, z_k) \cap [x_{g(k)-1}, x_{g(k)})|,$$

where $|\cdot|$ returns the number of integers in a set. We find the map that maximizes $S_g(\mathbf{Z}, \mathbf{X})$ and then define the metric (distance) between $\mathbf{Z}$ and $\mathbf{X}$ as $d(\mathbf{Z}, \mathbf{X}) = L - \max_g S_g(\mathbf{Z}, \mathbf{X})$. For example, if $L = 10$, $\mathbf{Z} = [3, 9]$ and $\mathbf{X} = 8$, then the desired map is $g^*(2) = 1$ and $g^*(3) = 2$, which maps segments 2 ($[3, 8]$) and 3 ($[9, 10]$) defined by $\mathbf{Z}$ to segments 1 ($[1, 7]$) and 2 ($[8, 10]$) defined by $\mathbf{X}$ respectively, and the resulting distance is 3. Note that the minimal distance between two distinct segmentations is 1, which implies a natural way to define connectedness. We say that a set $\mathbf{D}$ in this space is connected if for any two segmentations $\mathbf{Z}_a, \mathbf{Z}_b \in \mathbf{D}$, there exist $m$ segmentations $\mathbf{Z}_1, \ldots, \mathbf{Z}_m \in \mathbf{D}$ with $\mathbf{Z}_1 = \mathbf{Z}_a$ and $\mathbf{Z}_m = \mathbf{Z}_b$ such that $d(\mathbf{Z}_i, \mathbf{Z}_{i+1}) = 1$ for $i = 1, \ldots, m-1$.

Suppose $n+1$ segmentations have been sampled in a level set $C$ and the resulting NNDs in SLC are $d_1, d_2, \ldots, d_n$. Motivated by the observation that the histogram of $d_i$ ($i = 1, \ldots, n$) decays exponentially if $C$ is connected, we model them by a geometric distribution, $P(d_i|\beta) = \beta^{d_i}(1-\beta)$ ($d_i = 0, 1, \ldots$), where $\beta \in (0, 1)$ is an unknown parameter. We rank $d_i$ to obtain the order statistics $d_{(1)} \leq d_{(2)} \leq \cdots \leq d_{(n)}$. If $C$ consists of $K+1$ connected components, one expects the largest $K$ NNDs to be significantly greater than the rest when $n$ is large. By the similar missing data formulation as in Section 3.4, the MLE of $\beta$ if the largest $k$ distances are not observed is

$$\hat{\beta}_k = \frac{\sum_{i=1}^{n-k} d_{(i)} + k d_{(n-k)}}{\sum_{i=1}^{n-k} d_{(i)} + k d_{(n-k)} + (n-k)}.$$

Then the mean of the geometric distribution can be estimated by

$$\hat{\theta}_k = \frac{\hat{\beta}_k}{1-\hat{\beta}_k} = \frac{\sum_{i=1}^{n-k} d_{(i)} + kd_{(n-k)}}{n-k}.$$

From the memoryless property of the geometric distribution, the expected value of $d_i$ given $d_i \geq d_{(n-k)}$ and $\hat{\theta}_k$. Denote by $K_{\max}(=100)$ the predetermined maximal number of components. For $k = 1, \dots, K_{\max} - 1$ we compute $\hat{\theta}_k$ and define $\gamma_k = (d_{(n-k+1)} - d_{(n-k)})/\hat{\theta}_k$ as a statistic to test whether the observed largest $k$ NNDs are significantly greater than that expected from the other $n-k$ distances. Let $K_L \equiv 1$ and find $K_H = 1 + \max\{k \mid \gamma_k > \alpha, d_{(n-k+1)} \geq 2\}$, where $\alpha = 10$, and we define the maximum of an empty set as 0. Then the same pruning procedure as described for a continuous space by the size of a potential cluster is applied to obtain the final clusters of an empirical level set.

5.4. *A simulation study.* We simulated 10 DNA sequences each of length $L = 1000$ from the above segmentation model with four change points $[201, 401, 601, 801]$. Denote the nucleotide composition of the $i$th segment ($i = 1, \dots, 5$) by $\boldsymbol{\theta}_i = [\theta_{i1}, \dots, \theta_{i4}]$ for $a, c, g, t$, respectively. For the first four segments, $\theta_{ii} = 0.4$ and $\theta_{ij} = 0.2$ ($j \neq i$). For the last segment, $\theta_{5j} = 0.25$ for all $j$.

Set the maximal number of segments $N + 1 = L/100 = 10$. We found from a pilot study that, at $T = 0.5$, samples were concentrated around the global mode, while at $T = 2$ most samples were composed of 8 or 9 random change points. Thus, we applied the exact posterior sampling at 10 temperatures between 0.5 and 2 with geometric progression. For each temperature, 50K samples were generated. All the samples from different temperatures were partitioned into $M = 100$ level sets for estimating the tree of sublevel sets. The estimated trees have quite complicated structures, with an average of 45.3 local minima and 24.7 energy barriers. The number of local minima ranges from 21 to 74 and the number of energy barriers ranges from 6 to 49 for the 10 sequences. A local minimum in this discrete space is defined as a segmentation $\mathbf{Z}$ whose energy (10) is lower than the energy of all the neighbors (whose distance to $\mathbf{Z}$ equals one). For such a discrete space, it is feasible to verify an identified local minimum by this definition. Among the total of 453 detected minima on the 10 trees, 436 of them are true ones. We further applied steepest-descent optimization to the remaining 17 detected minima. It turned out that 16 of them led to distinct local minima on their respective trees and only one out of the 453 detected minima was produced by a false split of a terminal branch. These results demonstrated the high specificity of our tree construction algorithm.

We illustrate the results by one of the constructed trees. The posterior probabilities of the locations of change points ($T = 1$) are shown in Figure 6(A). As one can see, there are many peaks along the sequence and it is hard to find a reasonable principle to predict change points based on these marginal posterior probabilities, since the information on the combinatorial pattern among these potential change points is not revealed by such marginal statistics. However, the estimated tree of sublevel sets [Figure 6(B)] provides an informative way to understand this complicated posterior distribution. At the energy level of 1388, the sublevel set of the energy function of this posterior distribution is composed of three disconnected valleys. The first valley contains two local minima (indexed as 29 and 26 in the figure), the second valley contains a single local minimum 27, and the third valley contains many local minima. We report the respective lowest minima of the three valleys (minima 26, 27 and 6) in Table 3, from which we see that they are composed of 3, 5 and 4 change points, respectively. The change points of minimum 6, which is the global mode, are close to the four simulated ones. Minimum 26 does not contain the last change point, while minimum 27 splits the last change point into two. In addition, there exists another branch with a high-energy minimum 31, which has 3 change points around positions 400, 600 and 800. We further focus on the third valley which contains many local minima and report a few representative ones from different branches (minima 8, 12, 2 and 25) in the table. They all have four change points around the true locations, but with different combinations of local shift compared to minimum 6. The tree of sublevel sets definitely provided much more insights on the Bayesian inference of this problem: Not only did it detect multiple modes, but also recovered the hierarchy among them and revealed the combinatorial pattern among change points.

5.5. *Mouse upstream sequences.* A recent study identified eight genes that are up-regulated and function as activators in mouse embryonic stem cells [Zhou et al. (2007)], namely, Oct4, Sox2, Nanog, Esrrb, Tcf7, Nr5a2, Otx2 and Etv5. We extracted upstream 1,500 bases to downstream 500 bases around the transcription start sites and call them the upstream sequences of

TABLE 3

*Selected local minima of change points on a simulated sequence*

| Index | Change points | Index | Change points |
|---|---|---|---|
| 26 | [205, 393, 608] | 8 | [205, 393, 609, 787] |
| 27 | [205, 393, 608, 756, 816] | 12 | [205, 393, 609, 806] |
| 6 | [205, 393, 609, 813] | 2 | [199, 396, 609, 813] |
| 31 | [393, 609, 818] | 25 | [216, 394, 609, 813] |

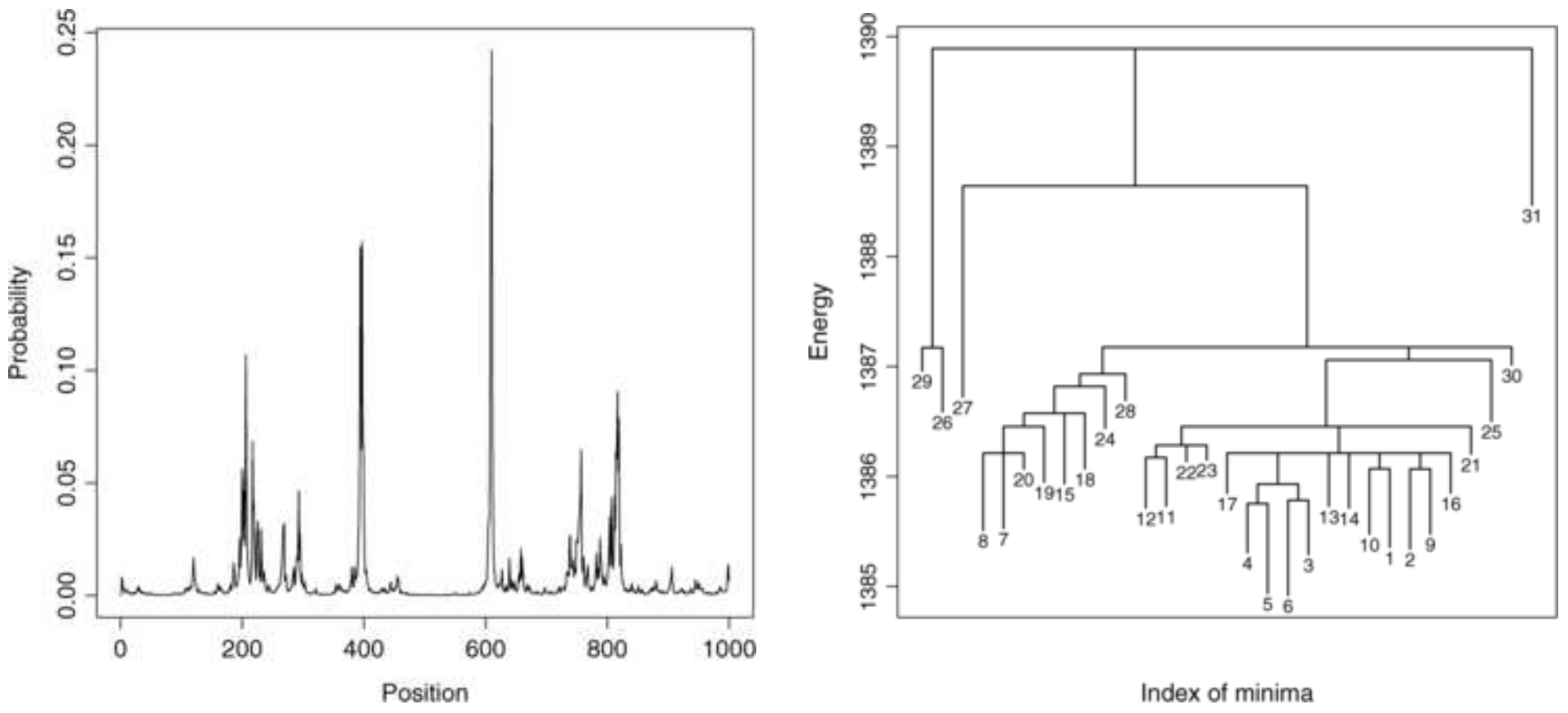

(A) (B)

Fig. 6. *Results for a simulated sequence.* (A) *Marginal posterior probabilities of change point locations along the sequence.* (B) *The estimated tree for the posterior distribution. The numbers in the figure are the indices of local minima.*

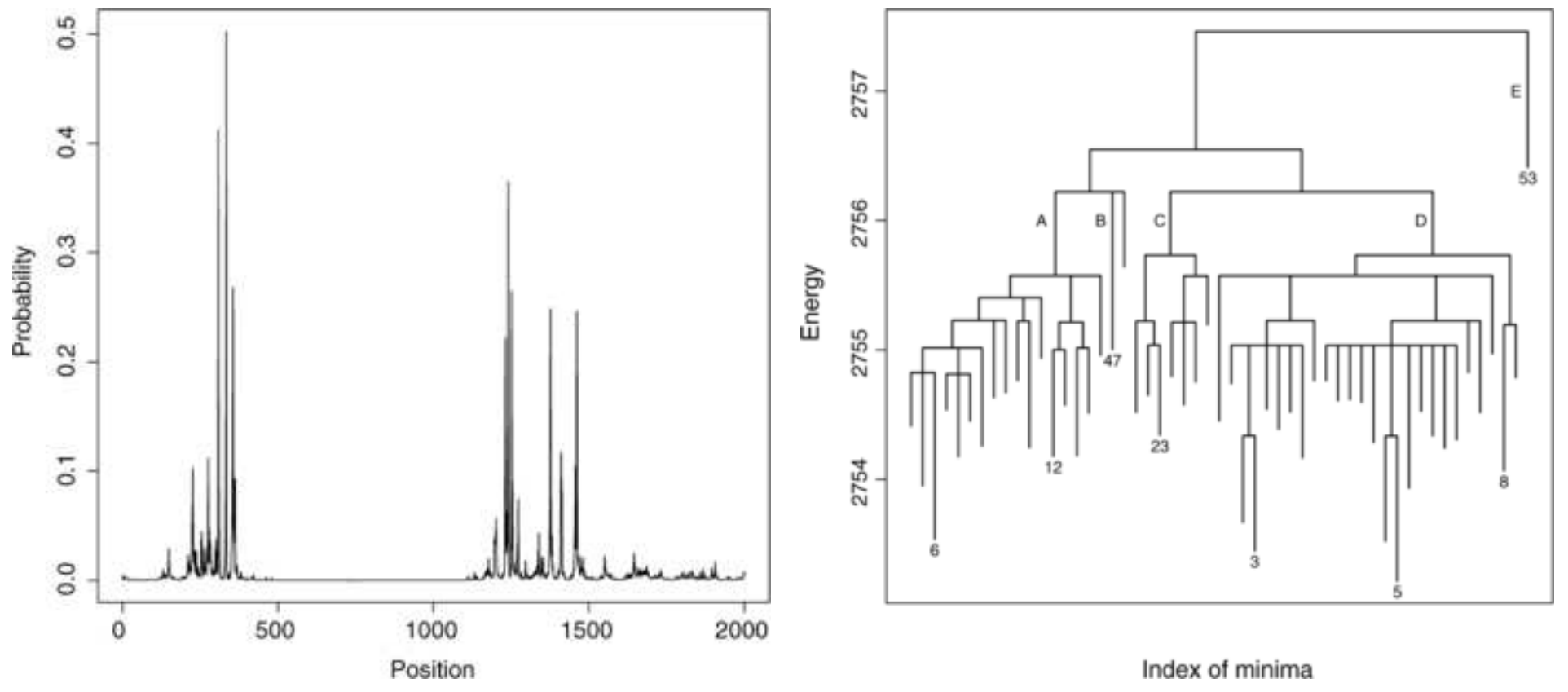

(A) (B)

Fig. 7. *Results for the upstream sequence of Nanog.* (A) *Marginal posterior probabilities of change point locations along the sequence.* (B) *The estimated tree of the posterior distribution with labeled valleys and minima.*

the genes. We set the maximal number of segments $N + 1 = L/100 = 20$ and applied posterior sampling followed by the BUP algorithm with exactly the same parameters as we used in the simulation study. The energy landscapes of the posterior distributions as revealed by estimated trees exhibit very different characteristics for different genes. The tree of the Oct4 upstream sequence contains only one minimum with no change points (i.e., the whole sequence is in one segment), while the tree of the Esrrb upstream contains

Table 4
*Selected local minima of change points on the Nanog sequence*

| **Valley** | **Minimum** | **Change points** |
|---|---|---|
| A | 6 | $[225, 361, 1376, 1460]$ |
| A | 12 | $[254, 361, 1376, 1460]$ |
| B | 47 | $[225, 361, 1376, 1410, 1461]$ |
| C | 23 | $[225, 307, 333, 355, 1376, 1410, 1461]$ |
| D | 3 | $[254, 307, 333, 355, 1376, 1460]$ |
| D | 5 | $[225, 307, 333, 355, 1376, 1460]$ |
| D | 8 | $[275, 307, 333, 355, 1376, 1460]$ |
| E | 53 | $[225, 307, 333, 355, 1240, 1251, 1376, 1413, 1460]$ |

55 local minima. On average there are 28.9 local minima and 17.6 barriers on the estimated trees. Among all the 231 detected minima on the eight trees, 227 of them are verified to be true ones, three of them lead to distinct minima via steepest-descent optimization, and only one of them corresponds to a false prediction.

We choose the results of the Nanog upstream sequence as an illustration. The posterior probabilities of change point locations are shown in Figure 7(A), where likely locations of change points indicated by peaks in the plot are mostly distributed in the intervals $[100, 400]$ and $[1100, 1500]$. (Note that the transcription start site is at position 1500.) The estimated tree of sublevel sets contains 53 local minima [Figure 7(B)] and a few deep energy valleys (big branches on the tree) labeled as $\mathrm{A}, \mathrm{B}, \ldots, \mathrm{E}$ in the figure. We select eight representative local minima labeled in the figure and reported in Table 4. The overall picture of the energy landscape is very clear. The deep valleys correspond to segmentations with different number of change points, such as valleys A and D which contain minima with 4 and 6 change points respectively. Within a valley defined in Figure 7(B), the local minima generally have similar combinations of change points. For example, the three labeled minima in valley D, each in a sub-branch of D, share five common change points, but differ in the location of the first one between 220 and 280 (Table 4).

Compared with the marginal posterior probabilities in Figure 7(A), the estimated tree clearly revealed much more information on the posterior distribution. From the marginal probabilities one can only identify change points from the local peaks with no information to determine the combination among them. However, the tree of sublevel sets not only identified different possible combinations of change points, but also organized them into a hierarchical structure that brings connectivity to differentiate and group these local minima. As illustrated, such information is very helpful for understanding the posterior distribution. One can view the posterior distribution as a multilevel mixture. At the first level, its energy landscape can be roughly represented by a mixture of a few large valleys [such as the ones labeled in Figure 7(B)]. Each valley may be further decomposed into smaller sub-valleys represented by various local minima.

## 6. Discussion.

We have formulated the tree of sublevel sets to characterize the energy landscape of a distribution, and developed the BUP algorithm to estimate the tree from Monte Carlo samples. The use of level sets as the building blocks in our method for tree construction has two advantages: (1) The samples in a level set are roughly uniform, which helps the development of algorithms to identify clusters; (2) Level set clustering requires much less computation as compared to clustering sublevel sets. The design of the BUP algorithm fits very well the EE sampler, which constructs empirical energy rings (level sets) for a wide spectrum. As we have mentioned in Section 3.5, a few other Monte Carlo methods are also good candidates for generating input samples with estimated density of states for the BUP algorithm.

Similar concepts of the tree have been used independently to describe the energy landscape of a physical system under the name of a disconnectivity graph [Becker and Karplus (1997)], with applications to peptide models [e.g., Krivov and Karplus (2002), Evans and Wales (2003)], lattice spin systems [e.g., Garstecki, Hoang and Cieplak (1999)] and protein folding pathways [Evans and Wales (2004), Carr and Wales (2005)] among others. A disconnectivity graph is constructed given a database of critical states of an energy surface, such as local minima, transition states (energy barriers) and pathways from a local minimum to a transition state. Optimization methods are often employed to search an energy surface for its critical states, mostly based on gradient approaches that utilize the first and second derivative matrices or ad hoc approximations for specific models [see Wales (2005) for a review]. The BUP algorithm in this article can also be used to construct the disconnectivity graph of a given potential energy surface. A unique feature of our method is that the construction of the tree is based on level set clustering of Monte Carlo samples and can be applied to any configuration space on which connectedness is defined through the use of a metric (or even a pseudo-metric). This is very important for statistical applications since a derivative-based search may be very difficult (e.g., for missing data problems) and even impossible (e.g., for discrete spaces). A firm comparison between the BUP algorithm and other chemical physics methods in constructing potential/free energy landscapes will be an interesting future direction of this work.

We believe that, with the fast development of powerful sampling methods and the exponential increase in computing capacity, computational statistical methods to extract useful information from large-size simulated samples, such as the BUP algorithm in this paper, are expected to play critical roles in many modern scientific fields.

## APPENDIX

PROOF OF LEMMA 1. Since $D$ is compact, for any $\varepsilon > 0$ its interior $D^0$ can be covered by a finite number of $\varepsilon$-balls, $B(\mathbf{x}_i, \varepsilon)$, where $\mathbf{x}_i \in D$ and $B_i = D^0 \cap B(\mathbf{x}_i, \varepsilon)$ is nonempty for $i = 1, \ldots, N$. Because $f(\mathbf{x}) > 0$ $\forall \mathbf{x} \in D$ and $f(\mathbf{x})$ is continuous, every $B_i$ has positive probability measure induced by $f$. The Harris recurrence of $\{\mathbf{X}_t\}$ implies that the chain will visit $B_i$ infinitely often (i.o.) with probability 1, that is, $P(\mathbf{X}_t \in B_i, \text{ i.o.}|\mathbf{X}_0 = \mathbf{x}) = 1$, for all $\mathbf{x} \in \mathcal{X}$ and all $i = 1, \ldots, N$. Since $N$ is finite,

$$P\left(\bigcap_{i=1}^{N} \{\mathbf{X}_t \in B_i, \text{ i.o.}\} \middle| \mathbf{X}_0 = \mathbf{x}\right) = 1.$$

Thus, there exists at least one $\mathbf{X}_t$ in each $B_i$ with probability 1 as $n \to \infty$, which implies that $\sup_{\mathbf{x} \in D^0} d(\mathbf{x}, \hat{D}_n) < 2\varepsilon$. Because $D^0$ is connected, any two points in $\hat{D}_n$ can be linked by a continuous path covered by a subset of $\{B_i\}$. Thus, the two points will be joined in SLC with maximal NND distance $< 4\varepsilon$, which completes the proof. □

PROOF OF LEMMA 3. (1) Let $V_i = \alpha_p r_i^p$ be the maximal volume of a ball centered at $\mathbf{X}_i$ that does not contain any other points, where $\alpha_p$ is a positive constant. Let $B_i$ be the open ball centered at $\mathbf{X}_i$ with volume $v/n$. The probability $P(V_i > v/n) = (1 - V_{B_i \cap C}/V_C)^{n-1}$, where $V_{B_i \cap C}$ is the volume of $B_i \cap C$. Since $h(\mathbf{x})$ is continuous, $P(B_i \subset C) \to 1$ as $n \to \infty$ and, thus, $P(V_i > v/n) \to (1 - v/(nV_C))^{n-1} \to e^{-v/V_C}$. This shows that the asymptotic distribution of $nr_i^p/V_C$ is identically exponential with mean $\theta = 1/\alpha_p$ for any $i$. Because $P(B_i \cap B_j \neq \phi) \to 0$ as $n \to \infty$, the joint probability $P(V_i > v/n, V_j > w/n) \to e^{-(v+w)/V_C}$ and, thus, $nr_i^p/V_C$ and $nr_j^p/V_C$ are independent asymptotically.

(2) We put $k$ balls of identical radius centered at $\mathbf{X}_{i_1}, \ldots, \mathbf{X}_{i_k}$. Then $\alpha_p d^p$ is the maximal ball volume such that none of the $k$ balls contains any points other than $\mathbf{X}^*$. Similarly, one can show that $P(\alpha_p d^p > v/n|\mathbf{X}^*) \to e^{-\beta v/V_C}$ as $n \to \infty$, where $1 < \beta \leq k$. Obviously, the conditional mean of $nd^p/V_C$ is $\theta/\beta$. □

**Acknowledgments.** We appreciate the editor, the associate editor and two referees for their critical comments and suggestions on an early version of this manuscript. We thank Yuan Yao for helpful discussions.

DEPARTMENT OF STATISTICS
UNIVERSITY OF CALIFORNIA
LOS ANGELES, CALIFORNIA 90095
USA
E-MAIL: zhou@stat.ucla.edu

DEPARTMENTS OF STATISTICS
AND HEALTH RESEARCH AND POLICY
STANFORD UNIVERSITY
STANFORD, CALIFORNIA 94305
USA
E-MAIL: whwong@stanford.edu